\documentclass[twocolumn,showpacs,preprintnumbers,pra,amsmath,amssymb]{revtex4}

\usepackage{graphicx}
\usepackage{bm}
\begin{document}
\title{Enhanced precision in entangled quantum clocks with phase estimation algorithm}

\author{Won-Young Hwang\footnote{Email: wyhwang@jnu.ac.kr}}

\affiliation{Department of Physics Education, Chonnam National University, Gwangju 61186, Republic of Korea
}
\begin{abstract}
We present an enhanced entangled-quantum-clock protocol that incorporates a quantum phase estimation algorithm to directly estimate proper-time differences as an unknown phase. By employing highly entangled multi-clock states, the achievable uncertainty scales inversely with the total number of quantum clocks, achieving the Heisenberg limit $O(1/N)$ while remaining completely immune to dynamical phase noise. Furthermore, we discuss the feasibility of utilizing this ultra-high precision protocol for gravitational wave detection, highlighting its potential for quantum-enhanced relativistic sensing and fundamental physics tests.
\end{abstract}
\maketitle
\section{Introduction}

Precise measurement of proper-time differences lies at the heart of both fundamental tests of relativity and emerging applications in quantum-enhanced metrology \cite{Hua24,Kan23,Yur86,Bol96,Gio04,Joz00,Bur01,Hwa02,Zyc11,Sin11,Bor25}. Since the pioneering proposal of entangled quantum clocks (EQCs) \cite{Hwa02}, quantum entanglement has been recognized as a qualitatively new resource for comparing proper times accumulated along distinct spacetime trajectories. In contrast to protocols based on two independent quantum clocks, the EQC protocol encodes the proper-time difference directly into a relative phase of an entangled two-qubit state. A key conceptual advantage of this approach is that the relevant phase remains fixed during the measurement process, thereby circumventing limitations associated with finite measurement durations and dynamical phase rotations.

Related and complementary aspects of quantum clocks and relativistic time effects have been explored from different perspectives. Interferometric visibility has been analyzed as an operational witness of proper-time differences, elucidating how relativistic effects manifest as coherence loss in quantum superpositions \cite{Zyc11,Sin11}. More recently, another variation of entangled quantum clocks has been proposed, which can test how proper-time difference due to gravity affects quantum superposition and interferences \cite{Bor25}.

In this work, we present an enhanced entangled quantum clock protocol that integrates a quantum-phase-estimation (QPE) algorithm \cite{Nie00} into the original EQC framework \cite{Hwa02}. We treat the accumulated relative phase as an unknown parameter to be estimated using systematic phase estimation techniques. In the meantime, the degree of entanglement between quantum clocks becomes higher, resulting in the Heisenberg-limited precision with well-known $\sqrt{N}$ advantages \cite{Yur86,Bol96,Gio04}.
While QPE has been extensively explored in quantum metrology to achieve Heisenberg-limited precision \cite{Hig07,Deg17,Gio11}, previous studies have predominantly focused on local phase parameters or single-frame frequency standards.

In the next section, we review the original EQC protocol. Then we present a protocol where QPE is integrated into the EQC protocol. In the protocol the phase value can be directly read out. We briefly discuss on how the $\sqrt{N}$ advantage can be obtained.

\section{Precision-enhanced EQCs with phase estimation algorithm}
In the EQC protocol, initially we prepare an entangled quantum clocks in a state
\begin{equation}
|\psi^{+}\rangle= |0\rangle_{A} |1\rangle_{B}+ |1\rangle_{A} |0\rangle_{B},
\label{1}
\end{equation}
where $A$ and $B$ respectively corresponds to each quantum clock whose proper-time will be compared. (The normalization factor is omitted almost in this paper.) Here $|p\rangle_{\alpha}$ denotes energy eigenstate with eigenvalues $E_p$ where $p=0,1$ and $\alpha= A,B$. We set $E_0= 0$ and $E_1= E$ without loss of generality. The time evolution of each quantum clock is in general given by a unitary operation $U_{\alpha}(t)|0\rangle_{\alpha}= |0\rangle_{\alpha}$ and $U_{\alpha}(t)|1\rangle_{\alpha}= e^{-iEt}|1\rangle_{\alpha}$, where $\hbar$ is set to be one. When two clocks follow different spacetime trajectories, the time for each clock is given by its own proper time. After proper-time $t_A$ and $t_B$ have elapsed for $A$ and $B$ quantum clocks, respectively, the initial state in Eq. (\ref{1}) becomes
\begin{eqnarray}
&& U_A(t_A)U_B(t_B)|\psi^{+}\rangle \nonumber\\
 &=& e^{-iEt_B}|0\rangle_{A} |1\rangle_{B}+ e^{-iEt_A} |1\rangle_{A} |0\rangle_{B} \nonumber\\
 &=& e^{-iEt_B}(|0\rangle_{A} |1\rangle_{B}+ e^{iE \Delta t} |1\rangle_{A} |0\rangle_{B}),
\label{2}
\end{eqnarray}
where the proper-time difference $t_B- t_A= \Delta t$.
Here, the information about the proper-time difference is encoded in the relative phase between the two states in Eq. (\ref{2}). The overall phase can be ignored.
In the protocol, we initially prepare quantum clock pair in the state $|\psi^{+}\rangle$ at a single site, as noted. Then we let each quantum clock depart and follows its own spacetime trajectory and subsequently gather them again. Then we perform a collective measurement that distinguish between $|\psi^{+}\rangle $ and $|\psi^{-}\rangle=  |0\rangle_{A} |1\rangle_{B}- |1\rangle_{A} |0\rangle_{B}$, which provides information about the relative phase and eventually the proper-time difference. In a protocol employing two separate quantum clocks, the phase of each clock rotates rapidly during the measurement, inevitably leading to dynamical averaging effects. In contrast, in the EQC protocol the relative phase to be measured is unchanged, so finiteness of measurement-time-interval $\delta t$ does not affect the measurement process \cite{Hwa02} (See FIG.1).
\begin{figure}
\includegraphics[trim=150pt 50pt 150pt 50pt, clip, width=\linewidth]{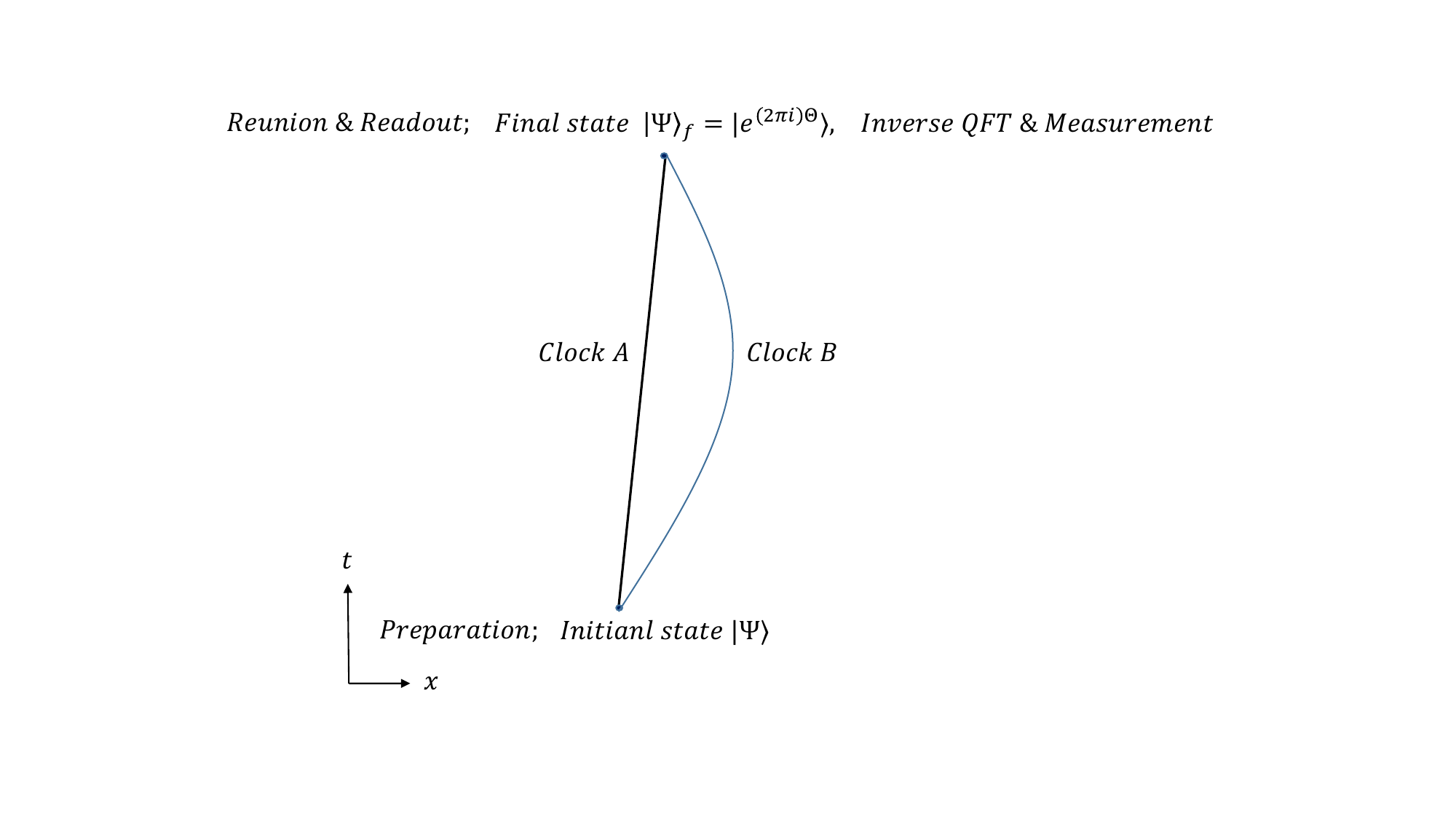}
\caption{Schematic spacetime diagram and operational workflow of the QPE-EQC protocol. The multi-clock entangled state $\vert{}\Psi\rangle$ in Eq. (\ref{3}) is initialized at a single site, after which Clocks A and B follow distinct spacetime trajectories. The proper-time difference $\Delta t$ is naturally encoded into the stationary relative phases of the entangled blocks during evolution. Upon reunion, the resulting state $\vert{}\Psi\rangle_f$ undergoes an inverse QFT followed by measurement, providing a direct digital readout of $\Delta t$.  }
\label{Fig-1}
\end{figure}

Despite this fundamental advantage, practical implementations of the EQC protocol still face challenges related to statistical uncertainty. In the original formulation, the achievable precision is ultimately constrained by projection noise and finite sample size. As a result, resolving small proper-time differences with high confidence typically requires a large number of entangled pairs. Moreover, even with $N$ pairs of quantum clocks, the achievable precision improves only by a factor of $\sqrt{N}$ due to statistical averaging, that is, the estimation uncertainty is proportional to $(1/\sqrt{N})$. However, in the protocol we propose the precision scales linearly with $N$. This is another case of the $\sqrt{N}$ improvements in quantum metrology \cite{Yur86,Bol96,Gio04}, which is possible due to high level of entanglement in the protocol.

We now present a precision-enhanced EQC protocol incorporating a quantum phase estimation algorithm.
Initially, we prepare a state
\begin{eqnarray}
|\Psi\rangle &=&
(|0\rangle_{A}^{2^{(n-1)}} |1\rangle_{B}^{2^{(n-1)}}+ |1\rangle_{A}^{2^{(n-1)}}|0\rangle_{B}^{2^{(n-1)}})
\nonumber\\
&&(|0\rangle_{A}^{2^{(n-2)}} |1\rangle_{B}^{2^{(n-2)}}+ |1\rangle_{A}^{2^{(n-2)}}|0\rangle_{B}^{2^{(n-2)}})\cdot \cdot \cdot   \nonumber\\
&&(|0\rangle_{A} |1\rangle_{B}+ |1\rangle_{A}|0\rangle_{B}),
\label{3}
\end{eqnarray}
where $|p\rangle^m= |p\rangle |p\rangle \cdot \cdot \cdot |p\rangle$, $m$ times product of $|p\rangle$ states and  $n,m$ are nonnegative integers.
Then we let quantum clocks depart and follow its own trajectory and subsequently gather them again. After proper-time $t_A$ and $t_B$ have elapsed for $A$ and $B$ quantum clocks, respectively, the initial state in Eq. (\ref{3}) becomes, ignoring overall phase,
\begin{eqnarray}
&&|\Psi\rangle_{f} \nonumber\\
&=&(|0\rangle_{A}^{2^{(n-1)}} |1\rangle_{B}^{2^{(n-1)}}+ e^{i2^{(n-1)}E \Delta t} |1\rangle_{A}^{2^{(n-1)}}|0\rangle_{B}^{2^{(n-1)}})
\nonumber\\
&&(|0\rangle_{A}^{2^{(n-2)}} |1\rangle_{B}^{2^{(n-2)}}+ e^{i2^{(n-2)}E \Delta t} |1\rangle_{A}^{2^{(n-2)}}|0\rangle_{B}^{2^{(n-2)}})\cdot \cdot \cdot   \nonumber\\
&&(|0\rangle_{A} |1\rangle_{B}+ e^{iE \Delta t} |1\rangle_{A}|0\rangle_{B}).
\label{4}
\end{eqnarray}
Let us introduce a state
\begin{eqnarray}
&&|e^{(2\pi i) \varphi}\rangle \nonumber\\
&\equiv&(|\tilde{0}\rangle+ e^{(2 \pi i) (2^{(n-1)}\varphi)} |\tilde{1}\rangle)(|\tilde{0}\rangle+ e^{(2 \pi i) (2^{(n-2)}\varphi)} |\tilde{1}\rangle)
\nonumber\\
&& \cdot \cdot \cdot
(|\tilde{0}\rangle+ e^{(2 \pi i) \varphi} |\tilde{1}\rangle),
\nonumber\\
\label{5}
\end{eqnarray}
where a phase $\varphi$ is estimated in the phase estimation algorithm \cite{Nie00}.
We treat $|0\rangle_{A}^m|1\rangle_{B}^m$ and $|1\rangle_{A}^m|0\rangle_{B}^m$ as $|\tilde{0}\rangle$ and $|\tilde{1}\rangle$, respectively, by restricting ourselves to the subspace spanned by these states. Then we can see that the state in Eq. (\ref{4}) can be written as
\begin{eqnarray}
|\Psi\rangle_{f}= |e^{(2\pi i) (\frac{E \Delta t}{2\pi})}\rangle,
\label{6}
\end{eqnarray}
where $\varphi$ is replaced by $(E \Delta t/2\pi) \equiv \Theta$.
Here the unknown phase $\Theta$ can be estimated as follows \cite{Nie00}. Here we take $N= 2^n$. We write the state $|j\rangle $ using binary representation $j=j_1 j_2 \ldots j_n$. Formally, $j= j_1 2^{n-1}+ j_2 2^{n-2}+ \cdot \cdot \cdot+ j_{n-1} 2^1+ j_{n} 2^0$. By quantum Fourier transform (QFT), the state $|j\rangle$ becomes
\begin{equation}
|\tilde{j}\rangle= \sum_{k=0}^{N-1} e^{(2\pi i)(k\frac{j}{N})} |k\rangle
= |e^{(2\pi i) (\frac{j}{N})}\rangle.
\label{7}
\end{equation}
Now the state $|\Psi\rangle_{f}$ can be decomposed as,
\begin{eqnarray}
|\Psi\rangle_{f}= \sum_{j=0}^{N-1} c_j |\tilde{j}\rangle.
\label{8}
\end{eqnarray}
Here the inner product
\begin{eqnarray}
c_j &=& (|\tilde{j}\rangle, |\Psi\rangle_{f}) \nonumber\\
&=& (|e^{(2\pi i) (\frac{j}{N})}\rangle, |e^{(2\pi i) \Theta}\rangle) \nonumber\\
&=& \sum_{k=0}^{N-1} e^{(2 \pi i)(\Theta- \frac{j}{N}) k}
\nonumber\\
&=& \sum_{k=0}^{N-1} z^k,
\label{9}
\end{eqnarray}
with $z\equiv e^{(2 \pi i)(\Theta- \frac{j}{N})}$. By performing inverse QFT on the state in Eq. (\ref{8}), we get
\begin{equation}
|\Psi\rangle_{f} \xrightarrow[QFT]{inverse} \sum_{j=0}^{N-1} c_j |j\rangle.
\label{10}
\end{equation}
Then we perform a measurement in the $|j\rangle$ basis. For a measurement outcome $m$, the optimal estimation for the $\Theta$ value is $(m/N)$. This optimal estimation method can be expected by observing that in the ideal case when $\Theta= (j^{\prime}/N)$ where $j^{\prime}$ is a nonnegative integer, all $c_{j}=0$ for $j \neq j^{\prime}$ and $c_{j}=1$ for  $j= j^{\prime}$.

Let us make a more rigorous analysis \cite{Nie00}.
First by Eq. (\ref{9}), recovering the omitted normalization constant, we get
\begin{eqnarray}
|c_j| &=& \frac{1}{N}\left| \frac{z^N-1}{z-1} \right|
=\frac{1}{N}\left| \frac{\bar{z}^N-1}{\bar{z}-1} \right|
\nonumber\\
&=& \frac{1}{N}\frac{|e^{(2 \pi i) N (\frac{j}{N}-\Theta)}-1|}{|e^{(2 \pi i) (\frac{j}{N}-\Theta)}-1|} \nonumber\\
&\leq& \frac{2}{N|e^{(2 \pi i) (\frac{j}{N}-\Theta)}-1|},
\label{11}
\end{eqnarray}
where $\bar{z}$ denotes complex conjugate of $z$ and we used an inequality $|e^{i \theta}- 1|\leq 2$.
Now consider an inequality $|e^{i \theta}- 1| \geq (2 | \theta |/ \pi)$ for $-\pi \leq \theta \leq \pi$. When $-(1/2) \leq (j/N- \Theta)\leq (1/2)$ the inequality can be directly applied to the last term in Eq. (\ref{11}), then we get a bound,
\begin{eqnarray}
|c_j|\leq \frac{1}{2|j- N \Theta|}.
\label{12}
\end{eqnarray}
If $(1/2)<(j/N- \Theta)$ then we add $-1$ in the term making it to be $|e^{(2 \pi i) (\frac{j}{N}-1-\Theta)}-1|$ without changing the value of term. Then we get a bound
\begin{eqnarray}
|c_j|\leq \frac{1}{2|j- N- N \Theta|}.
\label{13}
\end{eqnarray}
If $(j/N- \Theta)< -(1/2)$, similarly by adding $1$ we get 
\begin{eqnarray}
|c_j|\leq \frac{1}{2|j+ N- N \Theta|}.
\label{14}
\end{eqnarray}
Now we bound the probability of obtaining a measurement outcome $m$ such that $|m- N \Theta|> \Gamma$, where $\Gamma$ is a nonnegative integer characterising desired tolerance to error.
The probability is,
\begin{eqnarray}
&& p(|m-N\Theta|> \Gamma) \nonumber\\
&=& \sum_{0\leq j < N \Theta- \Gamma} |c_j|^2+
\sum_{N \Theta+ \Gamma < j \leq N} |c_j|^2 \nonumber\\
&\leq& \sum_{\Gamma \leq l }^{\frac{N}{2}}|\frac{1}{2 l}|^2+
\sum_{\Gamma \leq l}^{\frac{N}{2}}|\frac{1}{2 l}|^2 \nonumber\\
&\leq& \frac{1}{2} \int_{\Gamma-1}^{\infty} \frac{1}{l^2} \hspace{1mm} dl \nonumber\\
&=& \frac{1}{2(\Gamma-1)},
\label{15}
\end{eqnarray}
where we used inequalities in (\ref{12})-(\ref{14}) in getting first inequality.
Now we can obtain the value of $\Theta$ within uncertainty $\Delta \Theta= (\Gamma/N)$ with confidence probability $1-[1/2(\Gamma-1)]$. For example, with confidence probability $0.9$, if the measurement outcome is $m$ then the $\Theta$ value satisfies $m/N- 6/N \leq \Theta \leq m/N+ 6/N$. For a fixed confidence probability, $\Gamma$ is constant. Thus, for a fixed confidence probability, the uncertainty is proportional to  $(1/N)$, while total number of quantum clocks employed in the protocol is $2(N-1)$. Thus the uncertainty scales inversely with total number of quantum clocks.

Let us summarize the protocol. First prepare $N-1$ pairs of quantum clocks in the state $|\Psi\rangle$. After following two different trajectories in space-time, the state become $|\Psi\rangle_f$, which becomes $\sum_{j=0}^{N-1} c_j |j\rangle$ after inverse QFT is performed. Then a measurement in the $|j\rangle$ basis is done. For a measurement outcome $m$, the optimal estimation for $\Delta t$ is $(2\pi m/NE)$.

\section{Discussion and Conclusion}
To clearly highlight the performance of the proposed protocol, a quantitative theoretical comparison among different clock schemes is summarized in Table I.
\begin{table}[htbp]
\centering
\caption{Theoretical comparison among various quantum clock schemes and metrology protocols.}
\label{tab:comparison}
\begin{tabular}{p{0.22\linewidth} p{0.22\linewidth} p{0.18\linewidth} p{0.18\linewidth} p{0.20\linewidth}}
\hline\hline
\textbf{Protocol/ Scheme} & \textbf{Quantum State/ Resource} & \textbf{Precision Scaling} $\Delta(\Delta t)$ & \textbf{Dynamical Phase Noise} & \textbf{Readout Method} \\
\hline
Standard Unentangled Clocks & $N$ independent qubits ($|0\rangle + |1\rangle$) & $O(1/\sqrt{N})$ (SQL) & Affected (Rapid rotation during measurement) & Individual projection/Ramsey spectroscopy \\
Original EQC \cite{Hwa02} & $N/2$ pairs of $|\psi^+\rangle_{AB}$ & $O(1/\sqrt{N})$ (SQL) & \textbf{Immune} (Relative phase is stationary) & Collective Bell-state measurement ($|\psi^+\rangle$ vs $|\psi^-\rangle$) \\
Standard Quantum Metrology \cite{Bol96,Gio04} & Entangled probing state/$N$ channels & $O(1/N)$ (Heisenberg Limit) & Affected & Unitary phase readout/Projective measurement \\
\textbf{Proposed QPE-EQC (This Work)} & Multi-clock entangled state [Eq.(\ref{3})] & \textbf{$O(1/N)$ (Heisenberg Limit)} & \textbf{Immune} (Stationary relative phase) & \textbf{Direct Inverse QFT\& Bit-string readout} \\
\hline\hline
\end{tabular}
\end{table}
Standard independent clock synchronization schemes are constrained by the standard quantum limit $O(1/\sqrt{N})$ and suffer from dynamical phase rotation during finite measurement time intervals. The original EQC scheme circumvents the dynamical phase noise by encoding proper-time differences into a stationary relative phase, yet its precision remains bounded by $O(1/\sqrt{N})$. While general quantum metrology protocols \cite{Bol96,Gio04} achieve the Heisenberg scaling $O(1/N)$, they are affected by dynamical phase noise effect. Our QPE-enhanced EQC protocol uniquely combines the advantages of both frameworks: it completely eliminates dynamical phase errors inherent to relativistic time comparisons while simultaneously achieving the ultimate $O(1/N)$ Heisenberg precision scaling via a direct digital phase readout.

Regarding the computational overhead, our protocol exhibits exceptionally high resource efficiency in both state preparation and phase estimation. First, the preparation of the initial multi-clock entangled state in Eq. (3) entails generating independent GHZ-like blocks. Since the total number of physical qubits is $2(N-1)$, preparing this state requires only $O(N)$ CNOT operations, which is the absolute theoretical minimum for entangling $N$ qubits, with a highly parallelizable circuit depth of $O(\log_2 N)$. Second, during the phase estimation phase, standard QPE typically requires substantial circuit depth for controlled-$U^{2^k}$ operations. In our QPE-EQC framework, however, the relative phases proportional to $2^k E \Delta t$ are naturally accumulated through relativistic proper-time evolution along distinct trajectories, entirely bypassing the need for algorithmic controlled-unitary gates. The only additional algorithmic overhead is the final inverse QFT, requiring $O((\log_2 N)^2)$ elementary gates. Therefore, the protocol achieves the ultimate $O(1/N)$ Heisenberg precision at an essentially minimal gate complexity cost.

While a comprehensive analysis of gate errors and hardware-specific noise profiles in NISQ devices is beyond the scope of this conceptual proposal, our protocol provides the theoretical baseline for QPE-enhanced relativistic time measurement. Detailed numerical simulations incorporating specific hardware noise architectures are left for future work.

As noted above, previous studies of QPE in quantum metrology have predominantly focused on local phase parameters or single-frame frequency standards \cite{Hig07,Deg17,Gio11}.  Applying standard metrology protocols to spatially distributed clocks in a relativistic setting poses a fundamental challenge: the rapid dynamical phase rotation of independent clocks during the finite measurement interval introduces unavoidable averaging errors.
The true uniqueness of our QPE-EQC protocol is that it overcomes this fundamental limitation by bridging relativistic quantum information and algorithmic phase estimation. By encoding the proper-time difference directly into the stationary relative phase of a multi-clock entangled state in Eq.(\ref{3}), our scheme becomes completely immune to dynamical phase noise. Furthermore, unlike conventional QPE circuits that demand complex controlled-unitary operations, our framework utilizes the physical relativistic proper-time evolution itself to natively encode the $2^k$-scaled phases required for the algorithm. This demonstrates a qualitatively new synergy where relativistic kinematics act as the computational resource for optimal high-precision time comparison.

Compared to single-particle interferometric schemes (e.g., Zych et al. \cite{Zyc11})—where proper-time differences cause visibility decay and render practical implementations highly fragile against decoherence—the EQC framework \cite{Hwa02} avoids signal degradation by encoding $\Delta t$ directly into a stationary relative phase. This provides a far more robust and implementation-friendly platform for relativistic metrology. In addition, the multi-qubit EQC architecture opens up the unique possibility of leveraging quantum error correction  (QEC) \cite{Sho95,Ste96,Nie00} to protect temporal phase information. However, implementing QEC in quantum clock metrology presents an immediate challenge: logical qubit states ($\vert{}0\rangle_L$ and $\vert{}1\rangle_L$) generally do not possess constant, uniform energy eigenvalues, whereas physical time encoding strictly relies on well-defined energy splitting. Although resolving this energy inhomogeneity across logical states remains a nontrivial open problem, successfully integrating QEC with relativistic quantum clocks would represent a significant milestone toward fault-tolerant precision time metrology. (It is still also challenging to implement reliable quantum error correction, but recent development are hopeful \cite{Fow12,Kri22,Goo23,He25}.)

Integrating the proposed QPE-EQC framework with high-frequency metastable transitions establishes a theoretical ultimate precision that surpasses classical metrological boundaries. Governed by the Heisenberg-limited uncertainty scaling $\Delta(\Delta t)/T \sim \hbar/(E T N)$—where $E/\hbar = 2\pi \nu_0$ with $\nu_0$ being the clock transition frequency, $T$ is the operational duration, and $N$ is the number of entangled qubits—the relative precision can theoretically reach the $10^{-19}$ to $10^{-22}$ regime for multi-clock ensembles ($E/\hbar \sim 10^{15}\text{ Hz}$, $T \sim 10- 100\text{ s}$, and $N \sim 10^2 \text{--} 10^4$), and potentially down to $10^{-21} \text{--} 10^{-24}$ when extended to ultra-long-lived nuclear transitions (e.g., $^{229}\text{Th}$ \cite{Pei03,Sei19,Zha24}).
At this resolution, minute proper-time perturbations induced by millimeter- to micrometer-scale gravitational height variations or passing gravitational waves \cite{Bra71,Kol16} become directly resolvable, establishing the QPE-EQC protocol as a foundational architecture for quantum-enhanced relativistic sensing and fundamental physics tests.

In this work, we have proposed a precision-enhanced entangled quantum clock protocol by incorporating a quantum phase estimation algorithm into the original EQC framework. By interpreting the proper-time difference as an unknown phase parameter, the protocol enables a direct readout of relativistic time differences with Heisenberg-limited precision scaling $O(1/N)$ while remaining immune to dynamical phase noise. Furthermore, we discuss the feasibility of utilizing this ultra-high precision protocol for gravitational wave detection. These findings establish the QPE-EQC scheme as a promising architecture for quantum-enhanced relativistic sensing and fundamental physics tests.

\section*{Acknowledgement}
We are grateful to Prof. Kicheon Kang for insightful discussions.

\end{document}